\documentclass{article}

\usepackage{arxiv}

\usepackage[utf8]{inputenc} 
\usepackage[T1]{fontenc}    
\usepackage{hyperref}       
\usepackage{url}            
\usepackage{booktabs}       
\usepackage{amsfonts}       
\usepackage{nicefrac}       
\usepackage{microtype}      
\usepackage{lipsum}		
\usepackage{graphicx}
\usepackage{natbib}
\usepackage{doi}
\usepackage{epigraph}
\usepackage{makecell}

\usepackage[final,commandnameprefix=ifneeded]{changes}

\title{From risk to flourishing: Seeds for responding to a complex world}

\author{
    \href{https://orcid.org/0000-0002-9605-0007}{\includegraphics[scale=0.06]{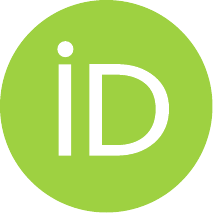}\hspace{1mm}Ryan M.~McGranaghan} \\
    Jet Propulsion Laboratory, California Institute of Technology\\
    Pasadena, CA 91109, USA\\
    \texttt{ryan.m.mcgranaghan@jpl.nasa.gov}
    \And
    \href{https://orcid.org/0000-0003-0630-3363}{\includegraphics[scale=0.06]{orcid.pdf}\hspace{1mm}Elisabeth H.~Krueger} \\
    Institute for Biodiversity and Ecosystem Dynamics\\
    University of Amsterdam\\
    Science Park 904, 1098 XH Amsterdam, Netherlands
    \And
    \href{https://orcid.org/0000-0002-8961-5292}{\includegraphics[scale=0.06]{orcid.pdf}\hspace{1mm}Alexandre Dunant} \\
    Eurac Research\\
    Viale Druso 1\\
    39100 Bolzano, Italy
    \And
    \href{https://orcid.org/0000-0002-6762-3937}{\includegraphics[scale=0.06]{orcid.pdf}\hspace{1mm}Guillaume Falmagne} \\
    Technological University Dublin\\
    Grangegorman Lower, Dublin 7, D07 H6K8, Ireland\\
    Trinity College Dublin\\
    College Green, Dublin 2, D02 PN40, Ireland
    \And
    \href{https://orcid.org/0000-0003-3356-3283}{\includegraphics[scale=0.06]{orcid.pdf}\hspace{1mm}Lyn\'ee L.~Turek-Hankins} \\
    Thayer School of Engineering\\
    Dartmouth College\\
    Hanover, NH 03755, USA
    \And
    \href{https://orcid.org/0000-0001-5533-5885}{\includegraphics[scale=0.06]{orcid.pdf}\hspace{1mm}Sara M.~Constantino} \\
    Doerr School of Sustainability\\
    Environmental Social Sciences\\
    Stanford University, Stanford, USA
    \And
    \href{https://orcid.org/0000-0003-1602-7243}{\includegraphics[scale=0.06]{orcid.pdf}\hspace{1mm}Raissa D'Souza} \\
    University of California, Davis\\
    Davis, CA 95616, USA\\
    Santa Fe Institute\\
    Santa Fe, NM 87501, USA
    \And
    \href{https://orcid.org/0000-0002-1678-3631}{\includegraphics[scale=0.06]{orcid.pdf}\hspace{1mm}Elke U.~Weber} \\
    School of Public and International Affairs\\
    Princeton University\\
    New Jersey, USA
    \And
    Graydon W.~Manzke \\
    Jet Propulsion Laboratory, California Institute of Technology\\
    Pasadena, CA 91109, USA
}

\renewcommand{\shorttitle}{From risk to flourishing}

\hypersetup{
pdftitle={From risk to flourishing: Seeds for responding to a complex world},
pdfsubject={physics.soc-ph},
pdfauthor={Ryan M. McGranaghan},
pdfkeywords={Complexity $|$  Risk $|$ Flourishing  $|$ Resilience $|$ Dialectic $|$ Social-Ecological-Technological Systems $|$ Deep Uncertainty $|$ Commons $|$ Care},
}

\begin{document}
\maketitle

\begin{abstract}

\added{Complex risk arises when interacting hazards, infrastructures, ecological systems, institutions, and human behaviors produce nonlinear impacts that cannot be inferred from any component in isolation. We call the study of these dynamics \textit{complex risk science}.}

\added{Flourishing is the actualization of all beings toward their good--not only individually but collectively--answering the question of `how do we live well with and for each other?'. In science as in society, the tendency to reduce, segregate, specialize, and settle has precluded embracing the concept of flourishing as a guiding principle. We describe how complex risk science reframes risk science research so that it can promote the flourishing of socio-ecological-technological systems (SETS), and explore how the practice of complex risk science may point society toward a more resilient future, more capable of flourishing.}

\added{Our exploration mirrors the unsettled and searching nature of flourishing itself: examine ideologies traditionally believed to be rigidly in opposition and consider what emerges in the space(s) and conversation between them. We call this a \textit{dialectical} approach: not a means of resolving contradiction, but a practice of sustaining it, so that systems, and our understanding of them, remain capable of reframing as conditions change.}

\added{Within this dialectical paradigm, grounded in conversation, pluralism, and pragmatism, we revisit three longstanding tensions in risk science: physical$|$social, determinism$|$uncertainty, and contextuality$|$generality. Although these tensions are not new, revisiting them together through the lens of complexity science reveals underexplored possibilities for the field--while taking seriously the institutional, intellectual, and economic constraints that shape risk science as it is actually practiced. Examining the conversation among these three tensions reveals five seeds that are dispositions for practicing complex risk science: interdependence, responsivity, participation, pluralism, and ongoingness. Indeed, these are not exclusive to risk science, nor should they be. They can support any scientific endeavor to embrace complexity and promote flourishing.}

\added{Practicing these seeds at scale, however, will depend on institutional reform: creating collectives toward complex risk science, supporting them with skillfully governed commons grounded in open science, and centering care in our research and evaluation.}
\end{abstract}


\section{Introduction}
This perspective aims to identify an approach to risk science research that \added{both tackles complexity and} promotes the flourishing of socio-ecological-technological systems (SETS). \added{Complex risk arises when hazards, human behaviors, and environmental, technological, social, and institutional systems interact non-linearly to produce
impacts that cannot be inferred from any component in isolation \citep{McPhearson2022, Krueger2025, Moore2024, Sillmann2024}. We call the study of these dynamics---how these systems \deleted{produce}\added{generate}, transmit, amplify, and adapt to risk across space and time---\textit{complex risk science}
\deleted{: a paradigm of risk science driven by complexity science, critical transitions, and resilience}.} 

Flourishing is the actualization of all beings toward their good---not only individually but collectively---answering the question of `how do we live well with and for each other?'. Despite this shared core\deleted{ of flourishing}, \added{different} notions of \deleted{its implications for society and how to reach it}\added{how to reach flourishing and its consequences on society} inevitably come into tension with one another, and require \deleted{an ongoing process of}\added{ongoing} \deleted{negotiation and} conversation, continuously exercised, continuously renegotiated. In science as in society, the tendency to reduce, segregate, specialize, and settle has precluded embracing \deleted{the concept of} flourishing as a guiding principle. While current crises have faltered trust in science as a basis for making sense of real-world processes\deleted{and for attaining flourishing for humans and non-human beings}
, they have \deleted{opened}\added{invited} the scientific community to \deleted{embrace}\added{adopt} other ways of creating knowledge and \deleted{of understanding our world}\added{understanding}.

We explore the normative role of \deleted{an orientation toward} flourishing \added{as a compass} for the field of risk science. In a context of \deleted{climbing}\added{rising} temperatures, \deleted{growing} social and economic inequality, \deleted{rising precarity of critical} infrastructure \added{fragility}, and persistent selection for the immediate over the long-term, the ways we have been doing and sharing science\added{(not limited to risk science)} have failed to grow--epistemologically and institutionally--to the immensity of risk in the 21st century. \added{Can the practice of (risk) science point} \deleted{Could (risk) science be done in a way that is capable of pointing} society toward a different future, one more resilient and more capable of flourishing? We \deleted{explore}\added{investigate} this possibility and \added{propose}\deleted{reveal} seeds \deleted{for ways of thinking and working}\added{for working} that could indeed shift science into such a \deleted{new} direction, building on \deleted{threads of thought from} various \added{intellectual traditions and} areas of knowledge creation. \added{Flourishing must be a conversation. And} our exploration mirrors \deleted{the conversational nature of flourishing itself: }\added{that by} consider\added{ing} ideologies traditionally held \deleted{believed to be rigidly} in opposition and \deleted{explore}\added{studying} what emerges in the space(s) and conversation between them. We call this a \textit{dialectical} approach.

Within this \textit{dialectical paradigm} grounded in \deleted{conversation,} pluralism\deleted{,} and pragmatism, we revisit three longstanding tensions in risk science: physical$|$social, determinism$|$uncertainty, and contextuality$|$generality. 

Although none of these are new, we contend that revisiting them together, through the lens of complexity science \added{and explicitly in light of flourishing as the normative commitment guiding this inquiry,}\deleted{and with explicit attention to the normative commitments that shape risk research,} reveals underexplored avenues for the field. Examining them as a joint set, each in conversation with the other, \deleted{allows us to explore the provisional seeds for what might be termed \textit{complex risk science}: interdependence, responsivity, participation, pluralism, openness, and ongoingness.} \added{allows us to identify five seeds for practicing complex risk science: interdependence, responsivity, participation, pluralism, and ongoingness.}

\added{Complex }risk science \added{will} enhance\added{s} our understanding of SETS and can help guide policy and actions to create resilient and flourishing communities and lives \citep{Beinhocker2023}. We conclude by examining how to bridge risk science practice and societal flourishing, a connection often \added{gestured at in adjacent literatures--sustainability science, economics, resilience thinking \citep{Folke2021, Beinhocker2023}--}\deleted{discussed }but \deleted{at best tenuously or superficially made for risk science specifically}\added{not, to our knowledge, substantively made within risk science itself}.

\added{Ultimately, }\deleted{E}\added{e}nacting the dialectical paradigm towards a risk science for a flourishing world will depend on \added{researchers adopting the seeds as guides for their work and institutions} creating collectives toward complex risk science, supporting them with skillfully governed commons, and centering care\added{--attention to vulnerability, and to the ongoing work of repair--}in \deleted{our}research and evaluation.

\section{The Modern Risk Landscape}


\added{Here, we motivate our approach with a snapshot of the challenges of modern risk science. 
The entanglement of systems \citep{Helbing2013} in today's world }\deleted{We live in a world of interdependent systems \citep{Helbing2013} whose entanglement} gives rise to \deleted{complex} risks and vulnerabilities that are difficult to quantify \citep{Vespignani2010, Cavalo2014, Danziger2016, Haimes2018, Reed2022, Dolan2025} and to deep uncertainties that challenge deterministic management paradigms \citep{Krueger_Ma_Kassab_Schulte-Romer_2025}. The interdependence of our 21st century world is not merely economic nor solely technological, but extends into the ecological and social realms\deleted{ as well}. 

In 1992, Ulrich Beck coined the term `risk society' to refer to the world of compounding threats we currently inhabit \citep{Beck1992}. 
Beck was thereby responding to the hazards and insecurities introduced by modernisation itself. The years since have proven his work prescient; \deleted{the potential for} compounding dynamics have proliferated in a world \deleted{dizzyingly} interwoven across people, goods, markets, information and ideas. \added{Modern risk is fundamentally embedded in complex social-ecological-technological systems (SETS) \citep{McPhearson2022}.}

This creates many of the benefits we enjoy but complicates the investigation of disruptions that can propagate across the planetary web. In most scientific disciplines, literatures and epistemological systems, awareness of these hazards of unprecedented proportions and uncertainties has grown, thereby highlighting the implications of interdependence across natural \deleted{systems}, technological \deleted{systems} \citep{Vespignani2010}, and social systems \citep{Helbing2013, Centeno2015}. \added{However, entangled decisions are typically taken independently, by separate entities}\deleted{The result is an entangling of decision-making across entities that typically act separately}~\citep{WorldEconomicForum2012, WorldEconomicForum2024}.
Approaches that isolate phenomena contributing to risk and \deleted{assume them to be}\added{treat them as} independent result in a miscalculation of their likelihood, coverage, and impact and a misu\deleted{nderstanding}\added{representation} of risk. Communities that rather acknowledge and work with interdependencies rather than artificially separate the constitutive elements of risk \added{are progressively forming to determine what practices and methodologies can}\deleted{have been trying to form to determine how they will work and what methodologies} meet the challenge. 

More than the fact of increasing connectedness \added{in a global risk landscape}, how things are connected matters. 
\deleted{Evidenced by the emergence of a global risk landscape from the planetary-scale interdependence of our systems and lives, many of the problems that we now face are wicked---being resistant}
\added{The planetary-scale interdependence of our systems and lives means problems are often ``wicked''---they resist} \deleted{to} any definitive "recipe-like" solution because they evolve, are interrelated to other problems, and \added{about which our knowledge is incomplete} \deleted{are defined by incomplete knowledge} \citep{Churchman1967, Ackoff1974, Rittel1973}. Intervention in wicked problems is fraught. Their entanglement with one another (interrelation) means that treating one creates issues in others \citep{Schofield_2024}. Their entanglement within (intrarelation) means feedbacks exist whereby interventions change the initial system state, therefore modifying its range of solutions. The engineering domain \deleted{may have well}\added{have} documented a perspective of interdependent systems and wicked problems (e.g., \citet{Madhavan2024}), but a science of risk demands sociological and anthropological perspectives, too \citep{Olsson2015}.

The deep uncertainties introduced by \added{interdependent} natural systems \deleted{interdependencies} overlaid with human dynamics cannot be easily studied through the classical risk science approaches \citep{AVEN20161}. Observed limitations in common risk science practices include: 
\begin{enumerate}
    \item Single-hazard orientations mischaracterize risk\added{, so} \deleted{and necessitates} modeling compounding and cascading interactions \added{is essential.} 
    \item Stationarity assumptions in historical risk models neglect the fact that complex systems are often non-ergodic (the space of possible states is only sparsely visited by past events so that past data is not predictive of the future) and \added{their environment evolves. This calls for }\deleted{ exist within changing environments, which require} nonstationary analyses and more dynamically updated risk models; 
    \item \deleted{Solely technological mitigation strategies neglect social, institutional, financial, and governance dynamics and reveal the role ethnography, sociology, psychology, and political economy must play in risk analyses.} \added{Technological solutions alone cannot explain or reduce complex risk. Disasters emerge through institutions, governance, economic incentives, public trust, community capacity, and human decision-making as much as through physical processes; the human dimensions of risk, not only of disaster, create path dependency, making the solutions we are able to draw on strongly dependent on what has happened in the past~\citep{Loorbach2017}. Technological innovation alone, consequently, \deleted{cannot adequately  characterize or reduce systemic}\added{is not sufficient to treat system-level} risk.}
\end{enumerate}

\added{These issues relate to how r}isk changes nonlinearly and sometimes non-intuitively due to complex effects of interdependencies. 
\deleted{Complex risk arises when interacting hazards, infrastructures, ecological systems, institutions, and human behaviors produce nonlinear impacts that cannot be inferred from any component in isolation.} 
%
%
\added{Additionally, d}isasters occur due to vulnerability, not hazards \citep{wisner2004risk}. Vulnerability is being altered as the frequency, magnitude, and location of extreme events change and increasingly intersect with one another \citep{Leichenko2008}. These changes are occurring more rapidly than our maps of them are being rewritten. The result is that unavoidable natural hazards tip over into avoidable disasters \citep{Stalhandske2024, Corburn2003}.

\added{We discussed above the limitations of risk science as a research practice--conceptual knowledge generated on the principles, methods, and models for understanding, assessing, and managing risk \citep{Antonsen2025}. Risk science as it is practiced is a different matter, and deserves to be characterized on its own terms before it is critiqued. In practice, the work of risk is not to encompass the greatest possible complexity, but to balance tradeoffs--constrained by institutional limits (resources, procedure, regulation), intellectual limits (the complexity any one team or process can actually bear), and cultural limits (the norms within which a decision becomes acceptable)--and it is, in the end, frequently an economic decision (and within an economics that struggles to value what is not immediately monetizable). This is also why a common actuarial criticism of risk science, that it rests only on historical data and is stationary or uni-modal, is somewhat overstated: efforts to characterize current and forward-looking risk are plentiful, but they struggle for identifiable reasons, including the difficulty of integrating continuous Earth observations with physical models, infrastructure dynamics, and changing societal conditions absent a stable baseline; the difficulty of incorporating hard-to-quantify or currently unobservable dimensions of risk, such as chronic risk or emerging `new normals'; and the underlying structures that produce systematic differences in vulnerability across groups \citep{SIMPSON2021489, PARSONS2026100770}. Nor is the decision-making risk science must serve confined to maps: it spans the public sector, the private sector, and civil society alike. Risk science, then, is not naive to its own complexity so much as constrained in how much of it any given practice can act on in a given moment--and studying that constraint is itself part of the field, in what might be called research \textit{on} risk science, alongside research \textit{in} it \citep{Antonsen2025}. This distinction between studying and practicing also anticipates a seed we return to below: it is one thing to study flourishing, and another to practice it--you cannot fully understand a system unless you change it \citep{Lewin1946, Schein1996, Quinn1999, Burns2026}, which is part of what participation, over hegemony, asks of the risk scientist.}

\section{Complexity \& Risk: Parallel Sciences}
\epigraph{The [21st] century will be the century of complexity.}{\textit{Stephen Hawking} \citep{Hawking2000}}

Long before globalization and technological revolutions shaped the final decade of the 20th century, complexity science was established in response to the realization that reducing the physical world to simple fundamental laws does not imply the ability to start from those laws and reconstruct the world~\citep{Anderson1972}. Complexity science is the study of phenomena that emerge from interactions of elements and processes; it is a paradigm of scientific discovery (e.g., \citet{McGranaghan2024}), rather than a ‘discipline’ \deleted{for} studying closed systems with steadfast boundaries and has produced a profound shift in scientific thought and concomitant understanding, perhaps most effective in its ability to evolve scientific disciplines toward new questions and framings.

For instance, \deleted{through complexity, notably }in systems ecology \citep{Shugart1979}, Darwin’s framing of competition as the driver of evolution gave way to ecology as the science of \deleted{interdependence, }networks, and emergent stability. Complexity science recasts ecology around resilience, feedbacks, and adaptive cycles, changing \added{its}\deleted{the field’s} central questions from ``Who outcompetes whom?'' to ``What structures and processes allow ecosystems to persist and adapt?''\deleted{--a question that requires continual re-examination}.

Complexity \deleted{is sometimes described as the boundary area}\added{can be understood as the boundary} between order and chaos~\citep{langton_computation_1990}. It is in this delicate space that scientific assumptions, often treated as opposing or irreconcilable categories, can come into conversation with each other, allowing for the continual re-examination \added{complexity requires}. We call this the dialectical approach: continually returning to \deleted{the} \added{a} conversation \added{in a way} that \deleted{opens new space where previously only a fixed boundary had been assumed}\added{transforms assumed fixed boundaries into new spaces of scientific practice}.

Due to its focus on interactions, feedbacks, and emergent phenomena, complexity science is at the forefront of risk and resilience research\deleted{ in the 21st century}~\citep{McPhearson2022, McGranaghan2024}. For instance\deleted{, in the physical sciences}, scientists have acknowledged the \added{physical} mutualities between ocean conditions, cloud formation, and temperature patterns across the globe \citep{NobelCommittee2021}, \added{but also}\deleted{and have been confronted with} the interconnectivity between economic, political, social, and biophysical systems \citep{Constantino2021}. Interconnections and feedbacks give rise to \deleted{the possibility of} non-linearities and \deleted{the emergence of} critical transitions\deleted{, which are realities that}\added{---and} we observe\added{ such transitions} across the planet \citep{Scheffer2009, lenton2025global}. The result is a\deleted{n altered} landscape of systemic risks, compounding and/or cascading changes across places and domains \citep{Simpson2025}. 
\added{The paradigm of c}omplexity \deleted{science is an essential paradigm offering a toolkit needed}\added{ offers a toolkit} to respond to our interdependent systems and lives and the risks that emerge from them.

\added{The chain this perspective traces from risk, to resilience, to flourishing begins with definitions from risk science itself.} Risk is often defined as the science that involves physical science (natural hazards) and their societal impact (exposure and vulnerability). These determinants of risk are consistent across the literature, \deleted{albeit with variations in their definitions}\added{although their definitions vary~}\citep{Simpson2021}. Vulnerability, the sensitivity of exposed critical infrastructure or populations to hazards, is \deleted{the flip side of}\added{intertwined with} adaptive capacity, \added{and it is their interplay that determines a population's or system's resilience}\deleted{while adaptive capacity is one of the key characteristics that makes populations resilient} \citep{Folke2002_AMBIO}. Resilience describes a system's \deleted{perseverance, its} ability to recover from and reorganize in response to disturbances, and the process of building adaptive capacity to \added{counteract vulnerability in order to} cope with unknown shocks and stressors~\citep{Folke2010_EcologySociety}. It is a process of seeking balance between maintaining and transforming. 

\deleted{Over the past several}\added{In recent} decades, both risk and resilience have matured from \added{a}disparate \added{set of observations into comprehensive frameworks for understanding: risk science now assesses inherently uncertain natural processes and their effects on socio-ecological systems, while resilience studies how those systems \textit{respond} to an increasingly wide range of shocks and disturbances} \deleted{descriptions of a set of phenomena toward frameworks for understanding inherently uncertain behaviors of the natural world (in risk science) and their effects in socio-ecological systems, while resilience accounts for the response of socio-ecological systems to a wider range of shocks or disturbances}
\citep{Folke_Colding_Berkes_2002, Krueger_2019_EarthFuture}. \deleted{Different}\added{Distinct} frameworks have grown in different fields such as disaster risk reduction \citep{Wisner2011, Sendai2015, UNDRR2025}, economics \citep{Hallegatte2014}, natural resource commons \citep{Ostrom2007, Ostrom2009}, climate \citep{IPCC2023}, \added{human agency and well-being \citep{BrownWestaway2011}}, and risk science itself \citep{Aven_2011, Burgess2016}.

\section{
\deleted{The Fact of SETS: Dialectic as Paradigm, Seeds as Practice} 
\added{Dialectic as a Paradigm for Flourishing SETS}} \label{dialectic section}
\epigraph{Our destiny is fated not only by great powers beyond our beckoning horizon but by the very way we shape and hold the everyday conversations of a familiar life.}{\textit{David Whyte} \citep{Whyte2015}}

\begin{figure}[tbhp]
\centering
\includegraphics[width=.8\linewidth]{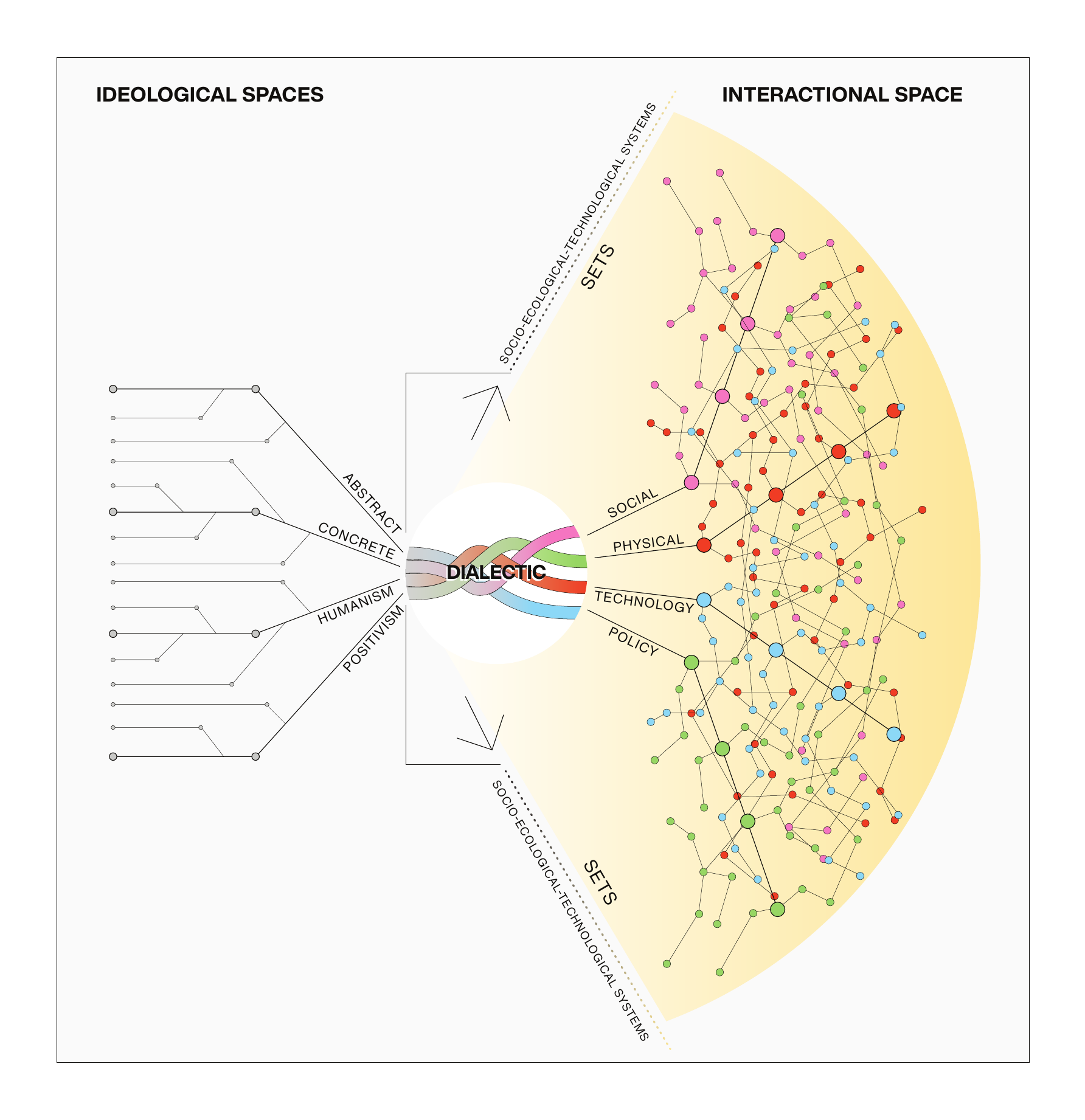}
\caption{Within the landscape of interconnected socio-ecological-technological systems (SETS), understanding and acting toward flourishing requires moving from rigid ideological spaces to a dynamic interactional space\deleted{;} \deleted{this move requires a paradigm of dialectic is required to affect this move}\added{, a move that requires a dialectical paradigm}. \\ (\textbf{Left, ideological spaces}) \deleted{Root ideologies of science--abstract, concrete, humanist, and positivist--represent distinct worldviews that artificially separate theory, material systems, values, and biophysical phenomena.}\added{Root axes of science, abstract--concrete and humanism--positivism, hardened into fixed positions, are what artificially separate theory from material systems, and values from biophysical phenomena.} These separate poles fragment understanding, creating tension and limiting engagement with the complexity of real-world systems.\\(\textbf{Right, interactional space}) SETS exist in a messy, entangled space where social, physical, technology, and policy dimensions are not new ideological positions but \deleted{are emergent} networks of ideas, practices, and structures \deleted{that emerge from merging}\added{that emerge from the dialectical interplay of} the perspectives on the left. \deleted{Dialectic is the way in which }We apprehend this interactional space \added{through dialectic}, \deleted{enabling dialogue,} weaving diverse knowledges~\added{\citep{Taylor2019}} and supporting pragmatic approaches that contribute to the flourishing of the scientific community as well as the human and more-than-human world.\\ 
\deleted{This framing recognizes that each dialectic is an indivisible duality, for which there can be no unambiguous resolution on either side of the binary \citep{Taylor2019}, but instead create the conditions}\deleted{The indivisible duality of each dialectic~\citep{Taylor2019} opens }\added{This opens space }for new insights, possibilities, and collective action within the complex realities of SETS.}
\label{fig:SETS_Dialectic}
\end{figure}

\deleted{We live in a deeply divided world with a rapidly changing risk landscape for which our existing tools are proving inadequate. The scientific tradition has created separate research directions in the natural, engineering and social sciences, economics, and the humanities. Within these disciplines we see further separation into disciplines and sub-disciplines, which can each embrace a real-world phenomenon, such as risk, but take distinct perspectives of how risk is defined, investigated, and interpreted. This pattern of separating the aspects of the problem has created a scientific tradition of risk science that reduces the problem into non-interacting components (e.g., physical, economic, psychological, social), whose lack of interaction obfuscate the similarities between them.} \added{The scientific tradition has long organized itself along disciplinary lines--natural, engineering, and social sciences, economics, the humanities--each \deleted{taking up}\added{studying} a shared phenomenon such as risk from a distinct vantage point, and with its own methods for defining, investigating, and interpreting it.} These separate perspectives harden into what could be called root ideologies of science: abstract, concrete, humanist, positivist (Figure \ref{fig:SETS_Dialectic}, left). \added{These separate perspectives harden into ideology along two axes: a \emph{formality} axis running from abstract to concrete, separating the formal sciences from the empirically grounded ones~\citep{Bunge1998}, and an \emph{epistemic} axis running from positivist to humanist, separating law-seeking, measurement-based inquiry from interpretive, meaning-based inquiry~\citep{Windelband1894}. Any given discipline sits somewhere on both (Figure \ref{fig:SETS_Dialectic}, left).}\added{Ideology, in this sense, functions like ``a straight-edge ruler in a fractal universe''\citep{Robinson2012}--a fixed measure held against a world that will not hold still.}

However, where we believe hard boundaries exist, interactions between the two sides of the boundary are possible and, like the ecotone between ecosystems, the spaces between the intellectual habitats become fertile crests \citep{Benyus1997}. The real world itself and the phenomena of scientific study do not know such separation or siloing, but exist as interdependent systems evolving from the interaction among biophysical, social, and cognitive-cultural processes across spatial and temporal scales.

\deleted{As an example of a cascading socio-ecological-technological failure that we will use throughout this perspective, consider the California Eaton and Palisades Fires in January 2025. Physically, these fires were a complex phenomenon emergent from an interdependent multi-hazard system (compounding wind and fire) \added{\citep{LosAngelesCounty2025}}. Culturally, they were the result of disagreement over which risk maps to use, and their devastation in part the result of trusting inaccurate estimates of wildfire risks in Altadena due to neglect of interactions between the hazards \citep{Haggerty2025}. Further complicating the preparedness picture was the availability of more complex wildfire risk models that could not be adopted due to either legal restrictions to adoption of closed-source essentially black-box models or trust, or both. Unfortunately, there are numerous and growing examples worldwide of natural hazards tipping into societal disasters as a result of inaccurate risk estimates, misaligned risk models, and failure to incorporate local understanding (e.g., \citep{Corburn2003}).}

\deleted{It is a fact that our world is constituted by Social-Ecological-Technological Systems (SETS)}\added{Most of our world is composed of SETS} and therefore by the interdependence of social (people, organization, laws, rules, norms, values, social relations, etc.), ecological (land, water, atmosphere, ecosystems, etc.), and technological elements (physical infrastructure and technologies) \citep{Walker2004, Markolf2018, Krueger2022, mcgranaghan2025entanglement}, and even simply the interdependence of physical infrastructure systems \citep{Rinaldi2001, Rourke2007, Danziger2016}. Key considerations for the resilience of these systems (both in terms of studying and governing) include \added{acknowledging}: 1) \deleted{acknowledging} the heterogeneity of SETS elements\added{, often}\deleted{ that often consist of a `bricolage' of} diverse and scattered \deleted{components} \citep{Frick-Trzebitzky2023}, 2) the connectivity of these systems across scales, giving rise to interacting slow and fast variables \citep{Biggs2012}, 3) \deleted{recognizing} who is involved, what their perspectives are, and how they influence the system at different scales \citep{peterson2024ethnographers, Krueger2025}. 
\deleted{To the tools of complexity science for studying emergent phenomena, SETS means we must add the}
\added{Therefore SETS, on top of complexity science tools, calls for }
methods for understanding contextuality, positionality and social construction \citep{Chambers2021, Klein2024}.

A paradigm is a framework of assumptions, principles, values, and methods \deleted{from which the members of the}\added{used by a} community \deleted{work} \citep{Kuhn1962, Pretorius2024}; a kind of generalization that characterizes the next stage of a community's work \citep{Anderson1972}. More important than a retelling of that history and the way in which paradigmatic thinking has led to the distinct core ideologies of science is understanding that these distinctions also create generative space between them.

SETS require \deleted{all of the ideologies traditionally associated with}\added{the root ideologies of} science to be in conversation\deleted{--boundaryless systems that call for disciplineless science}. ``Touch a limit of your understanding and it falls away, to reveal mystery upon mystery'' \citep{Robinson2015}. \deleted{Dialectic is the process by which we come to these limits, the places out of reach of our existing systems of thought, our disciplines and our ideologies. Flourishing means resisting hardening into ideology, an active process of seeking balance, attune to feedbacks. It works through holding dialectic, not resolving them.} 
\added{Dialectic, as we use the term here, is not a method for resolving contradiction into a higher synthesis. It is the practice of putting two things thought to be opposite, or irresolvable, into sustained relation, and attending to what emerges in the space between them rather than collapsing that space onto one side or the other. In Vermeer's famous \textit{Woman Holding a Balance}, a line of light divides the canvas; it is tempting to read the painting only in terms of what is illuminated and what is not. But the full painting is only apprehended when attention turns to the line itself--a space of substance and width, not a mere boundary. Dialectic performs this same turn of attention.} Reality, \deleted{on this view}\added{when addressed dialectically}, is never static but always in the process of becoming \citep{Whitehead1929, Deleuze1968, Stengers2018}\deleted{; the question d}\added{In d}ialectic \deleted{asks of} a system \added{does not have a fixed essence but is rather defined by}\deleted{is not \textit{what} it is but }\textit{when}, and in relation to what. \added{This is the same sense in which a system's resilience is never a property it has outright, but its distance from failure at a given moment, under given conditions. We return to this tension below in the Determinism$|$Uncertainty and Contextuality$|$Generality dialectics.} This is what allows dialectic to resist the hardening of position into ideology---a fixed, doctrinal certainty resistant to evidence--and instead keep systems, and the understanding we build of them, alive: capable of reframing as conditions change. 
``No self survives a real conversation'' \citep{Whyte2015}, and no ideology a dialectic.
When we hold the dialectics, rather than resolve them, we establish emergent networks of ideas, practices, and structures \deleted{composed of elements of}\added{originating from} the different ideologies. \deleted{Rather than}\added{In contrast to} ideologies, these networks \deleted{are held in entangled \deleted{spheres}\added{networks}:} \added{entangle} social, physical, technology, and policy \added{spheres} (Figure \ref{fig:SETS_Dialectic}, right).

Dialectic \deleted{becomes}\added{is where} \deleted{the transition from} the ideological spaces \added{transition} into the interactional space, the space of SETS. Emphasizing learning, and experimentation, the dialectic approach reveres plurality \deleted{and}\added{; it} requires conversation across all \deleted{relevant} actors concerned with an (open) system to understand \deleted{system}\added{its} dynamics and to identify adequate responses \citep{Biggs2012, Pahl-Wostl2009, Elmqvist2018}. We cannot expect a final quantification of the risk, but an ongoing engagement with it and \deleted{ways of studying and collaborating}\added{collaborative approaches} that support an ongoing process of assessment. Dialectic is how we apprehend SETS.

The structure of \added{the rest of} this paper mirrors the change we suggest for science: to embrace the conversational nature of the world \citep{Whyte2003}. Dialectic pairs, each an indivisible duality, are denoted \deleted{by `$|$' to create adjacency and interaction among them}\added{ with `$|$' to signify their adjacency and interaction}\deleted{, allowing for new understanding to emerge}. Dialectics are scaffolding for collective exploration\deleted{, together they are a framework that centers reflexivity and can}\added{; a reflexive framework--an awareness of the conditions shaping a response, and a critique of that response in relation to them \citep{Beck1992}--that} continually engages with changing conditions.
We explore the three key dialectics for risk science and detail what begins to emerge in the space(s) opened between them: physical$|$social, determinism$|$uncertainty, and contextuality$|$generality.

\deleted{The fact of change and need for an ongoing negotiation about the boundaries of the system have widened the definitions of}
\added{The behavior of SETS demonstrate that system boundaries and dynamics are constantly being negotiated. As a consequence, this pushed }risk and resilience to include transformation\deleted{, which underscores the importance of}\added{, moving beyond adaptive capacity toward rethinking the system itself}, \deleted{an}\added{both} element\added{s} missing from a more static risk science construction \citep{Park2012, Kates2012, Few2017, Folke2021}. \deleted{Thus, it is in adaptive capacity that risk and resilience come together.} However, we argue that risk science \deleted{must}\added{\textit{ought to}} go further\deleted{,}\added{ than incorporating resilience and transformation: it should} internalize a normative orientation toward \added{the good that defines} flourishing\deleted{ requires}.

\section{From Risk through Resilience to Flourishing}
\added{
\epigraph{Survival, mere continuation of being, is a condition man has in common with animals. Characteristic of humanity is concern for what to do with survival.}{\textit{Abraham Joshua Heschel} \citep{Heschel1996}}
}
The \deleted{starting point for}\added{central objective of} this perspective is to identify an approach to risk \deleted{science} research that promotes the flourishing of SETS. The flourishing of SETS is inextricable from the flourishing of society writ large, so this framing is as much about the well-being of individuals and communities as it is about systems.

We conceive of flourishing as the actualization of all beings toward their good. It is a continual process--one of negotiation and dialectic--between individual meaning and collective well-being created through our interactions with one another and emergent from them, all in response to the question, `How do we live well with and for each other?' Flourishing is a process, not a state, and will always be difficult to measure \citep{VanderWeele2025b}. However, it is the quality that a healthy society and sustainable systems must strive for. \added{Flourishing, understood this way, is not about resolution but about \textit{being among}--the same lesson that conversation, and dialectic, have to teach.}

\deleted{Understood as systems concepts, resilience and flourishing are undeniably related. However, that relationship has not been adequately described.} 
\deleted{Dialectic, as method, and flourishing, as aim, share a common shape: neither names a state to be reached but a practice to be sustained.} 
\added{Dialectic and flourishing are both a practice to be sustained rather than a state to be reached.}
Held together, dialectic, resilience, and flourishing form a single chain rather than three separate claims. Dialectic\deleted{, sustained rather than resolved,} is generative: resilience emerges from \added{it}\deleted{the process of dialectic}, as systems hold and rework their contradictions under changing conditions rather than allowing them to harden \citep{Woods2018}. Flourishing, in turn, is resilience directed normatively toward the good--not merely a system's capacity to go on, but a claim about what it goes on \textit{for}. A more actionable connection \deleted{becomes clear through a focus on}\added{ is revealed in} the `crisis discipline' of risk science, which exists at the nexus of complexity, sociology, collective intelligence, psychology, and political economy, and that we have identified is concerned with wicked problems. According to Michael Soul\'{e}, a crisis discipline is one that requires action before all information is known \citep{Soulé1985}\deleted{. They are those that}\added{; it must} cross `hard science' and methods of intuition, and, as \deleted{they}\added{it} pertains to existential problems, cross scientific research and socio-political action.

In risk science, resilience emphasizes process over state, agility over rigidity, responsiveness over prediction, adaptivity over reactivity\added{ \citep{Walker2004, Folke2010_EcologySociety, park2013, Woods2018}}. While resilience as a complex adaptive systems concept suggests the process of system response, flourishing suggests cultural work \citep{Moser2019}. Resilience perspectives have \added{already} expanded risk science's traditional focus on estimating the probability and consequences of adverse events to include system capacity to absorb and adapt to perturbations\added{ \citep{Peterson2003, Adams2016}}. Flourishing extends this trajectory by asking not only whether systems persist, but whether they enable individuals and communities to live well under conditions of continual change. Alone, resilience risks hardening into persistence or recovery in the face of shocks\deleted{ and disturbances; moving over and above it  toward} \added{; if risk science rather reaches for} the ideal of flourishing, \deleted{our risk science}\added{it} is called to more: the capacity to realize potential \citep{Nussbaum2011}.

For professor of philosophy Angie Hobbs, flourishing is about the actualization of potential, the fulfillment of all intellectual, imaginative, affective, and physical faculties; it is not and does not assume some eternal and final blissful state \added{\citep{Hobbs2018}}.
Flourishing, as resilience, is always unfinished, always relational. The \textit{Theory of Graceful Extensibility} proposes a requisite dynamic for all adaptive systems: they must change themselves to meet the \deleted{kinds of} challenges that arise as the world around them changes \citep{Woods2018}. This holds true for all systems whether that is a human body, planetary ecosystem, or a civilization. 
Just as resilience bridges \deleted{understanding of} complex adaptive systems \deleted{to ecological, social, political, and economic domains}\added{in ecology, society, politics and economics}, flourishing encompasses resilience and extends it\deleted{ to reflexivity--to thinking about the system in relation to its possibilities}\added{ to a commitment about what the better system is, which requires an examination of whose perception counts, for whom the system is to be better, and how to hold differing conceptions of `better' in relation to one another \citep{Durkheim2014, Haraway2014, WILLEN2025100377, Ingold2026}. This commitment that differentiates resilience from flourishing is itself a reflexive act, since it requires the system to continually turn its attention back on its own pursuit of the good.}

\deleted{The grand challenge in front of (risk) science is to understand and ensure the well-being of living beings, communities, and systems in the face of irreducible risks, understanding how they are embedded in a given context, and co-discovering solutions in ways that communities and their governments trust and find actionable. For this a normative orientation towards flourishing is required.}

\deleted{Sen and Nussbaum's founding move was to reject outcome-measurement (GDP, achieved states, utility) in favor of opportunity-measurement: not ``what did this person attain'' but ``what is this person able to do and be.'' The ``adaptive capacity'' list--mitigation, reserves, redundancy, replacement, operator flexibility, mutual aid, forecasting skill--is doing the identical move for systems rather than persons: not ``did the system fail'' but ``what reserves and degrees of freedom does it have before it would''. ``Distance from failure'' is, structurally, capability space. This is why ``not control, but conditions'' is the right implication to draw--control is an outcome-orientation (did we prevent the bad state), while capacity/capability is a conditions-orientation (does the system, or the person, have room to maneuver).}

\deleted{Conceptions of Flourishing is inherently practical: it concerns the real capabilities of individuals and communities rather than abstract ideals, as emphasized in the \textit{Capabilities Approach} developed by Martha Nussbaum \citep{Nussbaum2011}. It is, in this sense, a capacity; not only to live well individually, but to collectively reclaim the public life and spaces we share \citep{Allen2023}.
Attending to flourishing therefore exposes the tension within risk science between knowledge produced to explain the world and that needed to act upon it.} 
\added{The \textit{Capabilities Approach}, developed by Amartya Sen and Martha Nussbaum, distinguishes capabilities from functionings to honor \textit{choice} as a way to allow different persons to work toward the good they perceive \citep{Nussbaum2011}. Adaptive capacity effects the same change for resilience, evidenced by its own vocabulary: mitigation, reserves, redundancy, replacement, operator flexibility, mutual aid, forecasting skill \citep{bruneau2003, Hollnagel2006}. Read this way, ``not control, but conditions'' follows directly: control asks whether a single outcome was secured, while capacity and capability alike ask whether a system or a person retains room to maneuver toward whichever conception of the good it is called to pursue. Flourishing is inherently practical in this sense: it concerns the real capabilities of individuals and communities to live well under the conditions they face, rather than abstract ideals alone, and extends beyond the individual to collectively reclaim the public life and spaces we share \citep{Allen2023}. Flourishing's practicality exposes a tension already latent in risk science, between knowledge produced to explain the world and that needed to act upon it.}

\begin{figure}[tbhp]
\centering
\includegraphics[width=.8\linewidth]{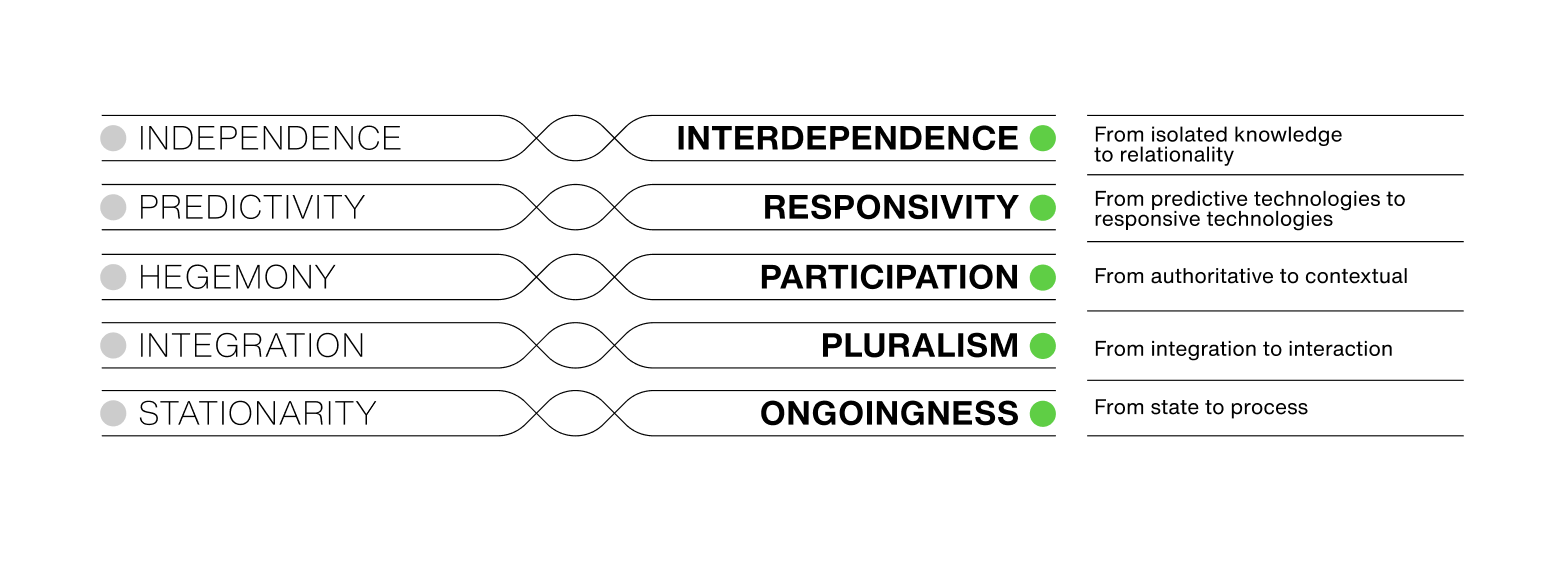}
\caption{This perspective adopts flourishing as a normative frame and dialectic as \deleted{the}\added{a} paradigm. Together, they suggest \deleted{six}\added{five} seeds, each a disposition of inquiry and daily activity that can enact the paradigm in risk science research and practice.}
\label{fig:six_principles}
\end{figure}

\deleted{In this way, an orientation toward flourishing reveals a}\added{This} dialectic between theory and practice \deleted{that} arches across the \added{three dialectic pairs we discuss in the next section} \deleted{other tensions discussed here}. \deleted{Therefore,} We suggest \deleted{six}\added{five} seeds that \deleted{help move inquiry and action around risk} \added{bring risk practices} toward the dialectical paradigm and therefore, toward flourishing \deleted{(\ref{fig:six_principles})}. \added{These seeds are dispositions of a researcher. Discerned by examining the research that engages well with three tensions within risk science, they are not particular to risk science, nor should they be: these weightings of attention and judgment, in support of a disposition toward flourishing, are strengthened by their value both within and beyond risk science. They are not independent of the authors' subjectivities, but as supports for orienting toward flourishing, by our own definition, they should not be.}

\deleted{These seeds (Fig. \ref{fig:six_principles}, middle) emphasize feedbacks necessary to establish systems change (from left to right) toward a dialectical paradigm \citep{Varela1991}.}
\added{This shift is represented from left to right in Fig. \ref{fig:six_principles}, culminating in the suggested seeds on the right; they emphasize the feedbacks that support this shift and sustain new practices in their movement toward systemic change~\citep{Varela1991}.}


\deleted{What we are suggesting}\added{ Our proposed seeds}
is substantiated by a nexus of complexity science and sociological and anthropological practices that center positionality and contextuality. This includes both qualitative and quantitative approaches, ethnographic engagement, and a praxis of intimacy, inquiry, and conversation \citep{Davis2016, Norstrom2020, Caniglia2021}. We argue that a lens of flourishing will guide the risk science community toward a paradigm of dialectic and plurality\deleted{, and conversation}, one capable of \deleted{supporting science and collective action in the face of} \added{ handling the }complexity and deep uncertainty\deleted{, key elements} of socio-ecological-technological (i.e., `wicked' \citep{Rittel1973, Madhavan2024}) problems. 

\section{Dialectics: The \deleted{Advent}\added{Practice} of Complex Risk Science}

We discuss three central \deleted{dialectics}\added{tensions} in \deleted{complex}risk science that must be held and put into conversation\added{ to move toward complex risk science, and indeed its contours are already visible in the arenas of c}\deleted{. C}limate risk science and ecology \deleted{have become arenas where the contours of a complex risk science are already visible}. We use the example of cascading wildfire risk involving electrical infrastructure and extreme wind conditions in Southern California during the Eaton and Palisades fires as a recurring illustration of how these \deleted{dialectics}\added{tensions} manifest in practice and additionally how \added{exploring the conversation within them} they can support research and action more capable of creating flourishing\added{ for SETS}.

\subsection*{Physical$|$Social: \deleted{Implication from the social}\added{ The human} side of risk}

Different communities interpret the concept of ‘risk’ in different ways. For the general public, risk often refers to adverse consequences, worrying about the ‘risk of rain’ for an outdoor wedding or the ‘risk of frost ruining a citrus crop’. Economics, statistics, and engineering use the term to refer to the unpredictability of outcomes, typically quantified as the variance of possible outcomes of an event or action around the expected value. Psychology, in contrast, treats risk as a subjective construct that may differ across individuals, groups, cultures, and situations \citep{Loewenstein2001}. It involves not only probabilities and outcomes but also emotions (e.g., feelings of dread, vulnerability, catastrophic potential, and uncontrollability \citep{Loewenstein2001}). Opposing a purely quantified conceptualization, Douglas and Wildavsky \citep{Douglas1983} argued that risks are hidden, selected, contested, and biased. Risk perception is a subjective impression, influenced by cultural and contextual factors, and this surprising discovery remains foundational to risk science \citep{Loewenstein2001}. Risk always points to the future. It is therefore a product of imagination and, by extension, of culture \citep{Adams2016}. Anthropological accounts show that risks are created and interpreted within social systems: their magnitude depends on the quality of social relations, the strength of institutions, and the cultural processes that render individuals and communities legible or illegible \citep{Beck1992}. James C. Scott \citep{Scott1998} argued that states simplify complex systems to facilitate control and resource allocation. Scott’s ‘legibility’ can be grounded in disaster policy: U.S. federal flood insurance maps reduce lived complexity to zones of “in” or “out,” yet local communities experience flood risk as gradational, contextual, and embedded in social ties. 

The January 2025 California Eaton and Palisades fires, too, illustrate how risk manifests through inseparable physical and social dynamics. Preliminary investigations linking ignition to electrical distribution failures highlight the physical infrastructure dimension~\added{\citep{LosAngelesCounty2025}}, while patterns of exposure, preparedness, insurance access, and evacuation capacity, queued on a wildfire risk model that assumed independence between wildfire and extreme wind, were ill-matched to conditions \added{in communities such as Altadena~\citep{Haggerty2025}}.
Further, inadequate granularity of the social context in these areas in the aftermath of the disaster reveal deeply social structures of vulnerability. Treating wildfire risk primarily as a biophysical hazard obscures how infrastructure governance, land-use decisions, and community-level adaptive capacity jointly shape hazard preparedness and disaster outcomes.

These simplifications expose \added{the consequences of} simplified risk quantification\deleted{in action, translating}\added{, when} lived uncertainty \added{is translated }into standardized categories. Risk perception also depends on trust--in people, organizations, and institutions--which in turn shapes attitudes, behaviors, and the governance of risk \citep{Wirz2025}. With trust come cascading considerations, from the affective nature of risk to the ways information is shared and acted upon \citep{Loewenstein2001}. The tension between legibility and complexity remains (and is finally irresolvable). To hold the complexity with integrity requires \deleted{a plural and}\added{ our} dialectical approach\deleted{: one that explores}\added{, continuously engaging with} the interstitial space between the physical and social dimensions and acknowledging their interactions\deleted{ and the need for ongoing engagement and reconsideration}. In the absence of capable estimates, renewed on an appropriate timescale and spatial resolution, we are unable to develop an orientation (social and perceptual) toward risk. These relationships are not simple. For instance, from the climate domain, psychological distance--the sense of how far away a risk or event feels--has often been seen as a barrier to climate action \citep{Singh2017}. \deleted{But research shows mixed results, questioning the}\added{ However, results are mixed, weakening this} idea that greater distance always reduces people’s willingness to \deleted{support or} take action on climate change \citep{vanValkengoed2023}. 

Therefore, climate risk science increasingly draws on social sciences, from cultural cognition to participatory mapping, to negotiate the tension between abstract, global framings and lived, local experiences of risk (cf. the Yale Program on Climate Change Communication “Climate Change in the American Mind (CCAM) Explorer”) \citep{Ballew2019, Yale2025}. Embracing this holistic conceptualization and investigation of risk requires more than integration, but a dialectical grafting that recognizes the distinction between complexity science, social science, and humanistic sciences and creates interaction between them so that something new emerges \citep{Keller2021}. Vulnerability, a constitutive element of risk, \deleted{forms the bridge between}\added{ bridges} the physical and the social. It is a multidimensional construct shaped by structural, economic, institutional, and social factors, and it demands both quantitative and qualitative methods \citep{Fuchs2012}. Vulnerability is emergent, involving feedbacks and nonlinearity. For example, Hurricane Katrina revealed vulnerability as both physical (levee failures, storm surge) and social (race, poverty, institutional neglect): the disaster’s impact cannot be understood without the entanglement of both dimensions. Risk’s cultural components must be joined with the physical and the quantifiable in guiding how we think about, enact, and communicate risk science. 
Beck critiqued the cultural-political hegemony of scientism, which has instrumentalized \deleted{and reduced} risk, and called for modernization to become reflexive: self-referential and self-critiquing \citep{Beck1992}. \added{Beck's call is similar to one that risk-governance scholarship has since carried directly into practice, arguing that managing complex risk requires institutions themselves to become objects of ongoing scrutiny and revision \citep{Renn2008, Voss2006}.}

\subsection*{Determinism$|$Uncertainty: Toward a science of complex risk}
\epigraph{In time, those Unconscionable Maps no longer satisfied, and the Cartographers Guilds struck a Map of the Empire whose size was that of the Empire, and which coincided point for point with it. The following Generations, who were not so fond of the Study of Cartography as their Forebears had been, saw that that vast map was Useless, and not without some Pitilessness was it, that they delivered it up to the Inclemencies of Sun and Winters.} {\textit{Jorge Luis Borge} \citep{Borges1972}}

\deleted{\textit{Can we really model complex risk?}}

Complex risk emerges from layered interdependencies between the biophysical and the socio-economic \citep{Simpson2021}. These risks exist within regimes of deep uncertainty, without predictable probability distributions, and systemic entanglement, where system boundaries are ill-defined. Complex risk calculations require acknowledgement of deep unknowns. Deep uncertainty arises when analysts or decision makers cannot agree upon the models or probability distributions by which outcomes should be judged \citep{Lempert2003}. Table 1, adapted from Walker et al. \citep{Walker2013}, shows the gradations of uncertainty and their practical implications.

\begin{table}[t!]
\centering
\caption{Types of uncertainty and implications for analysis (adapted from \citet{Walker2013})}
\begin{tabular}{p{0.28\linewidth}p{0.3\linewidth}p{0.3\linewidth}}
\toprule
\textbf{Uncertainty Level} & \textbf{Description} & \textbf{Potential Approach} \\
\midrule
Complete Certainty & Outcomes are fully known and predictable. & Deterministic models; optimization techniques. \\
\makecell{Level 1: \newline Low Uncertainty} & Minor uncertainties exist; outcomes are mostly predictable. & Sensitivity analysis; scenario planning with minor variations; statistical uncertainties from limited past data. \\
\makecell{Level 2: \newline Moderate\newline Uncertainty} & Probabilities can be assigned to outcomes based on historical data. & Probabilistic risk assessment; stochastic modeling; systematic uncertainties and model variations \\
\makecell{Level 3: \newline High Uncertainty} & Multiple plausible outcomes exist; probabilities are difficult to assign. & Scenario analysis; continuous interactions between field observations and models; exploratory modeling. \\
\makecell{Level 4: \newline Deep Uncertainty} & Fundamental disagreements or unknowns about system behavior and outcomes. & Adaptive policymaking; dynamic adaptive policy pathways; resilience planning. \\
\makecell{Level 5: \newline Total Ignorance} & Complete lack of knowledge about potential outcomes or system behavior. & Precautionary principle; building flexibility and adaptability into systems and policies. \\
\bottomrule
\end{tabular}
\end{table}

The levels above refer to epistemic uncertainty, which invites changes in our approach to risk science. But complex risks also confront us with ontological uncertainty: cases where the system and its boundaries evolve, the data on which models were tuned become irrelevant, and feedbacks, tipping points, and cascading effects undermine the very basis of extrapolation. The standard remedies for moderate uncertainty levels are to vary the model parameters or even model types, resulting in systematic or theoretical uncertainties beyond the well-understood statistical uncertainties; but these extensions fail in the presence of deep unknowns.

Modern risk science has traditionally relied on ergodicity, the assumption that most possible configurations of the system have been explored in historical observations, enabling probabilistic modelling and long-range forecasting. Yet many SETS are non-ergodic: they do not revisit prior states, and the future is not a variation on the past but a domain of expanding possibility shaped by emergent dynamics \citep{Varela1999}. These systems exhibit open-ended novelty, where the space of possible outcomes itself is under construction rather than a closed frame for estimating probabilities from past observations. Traditional methods fail not just technically, but philosophically.

Returning to the Eaton and Palisades fires example, wildfire risk models increasingly incorporate high-resolution environmental predictors, yet cascading processes (e.g., infrastructure-triggered ignitions under extreme wind conditions) remain difficult to represent deterministically. Those events illustrate how uncertainty is not merely a function of incomplete data but of structural limits in modeling interacting hazards and, further, socio-technical responses. Moreover, when advanced models are inaccessible due to proprietary constraints or lack of interpretability, epistemic uncertainty becomes institutional uncertainty~\added{\citep{Haggerty2025}}, shaping whether knowledge can meaningfully inform action.

SETS require a new epistemology, one that embraces open-endedness and the co-evolution of risk and response \citep{Hollnagel2006, AVEN20161}. Beyond what Krakauer \deleted{[2023]} \citep{Krakauer2023} terms the Feynman Limit--the boundary where formal models lose descriptive and predictive power--we must, therefore, rely on generative, abstract structures that synthesize high-dimensional inputs into forms that can guide decision-making under deep uncertainty \citep{Varela1999}. Practically, this suggests complementary analysis modes, which might be likened to position and momentum spaces in physics \citep{Hall2013}. Positional Analysis precisely defines causally stable system states: "What is the threat's nature and magnitude now?". Momentum Analysis understands dynamics shaping causally open systems over time: "How do we build adaptive capacity?". In practice, these are inseparable; a flood's position entangles water height with social vulnerability and its momentum creates a feedback of economic activity with local climate and social adaptation. The complementarity of these analysis modes underscores our thesis: adopting a dialectical, pluralistic paradigm is not about resolving binaries, but exploring the generative conversation between them.

These needs are best illustrated in climate science. Early Intergovernmental Panel on Climate Change (IPCC) reports framed the problem largely in terms of climate impacts. By the Fifth Assessment (AR5), risk was explicitly adopted as the central frame: hazards, exposures, and vulnerabilities combine to produce complex risks \citep{IPCC2012}. By AR6, the focus shifted further, to dilemmas of adaptation versus mitigation under cascading, interacting hazards. ‘Climate risk science’ has then been shaped by the tension between eliminating risk through mitigation and living with risk through adaptation. Mitigation and adaptation suggest radically different responses. Traditional risk analysis assumes stable distributions, return periods, and separable hazards. Conversely, climate change issues are nonstationary, interactive, and deeply uncertain. Climate change is a ‘meta-crisis’: many interacting sources of uncertainty in varied societal decisions, juxtaposed with cognitive and motivational biases (both in the general public and among scientists and policymakers) that exaggerate our perceived ability to predict \citep{Loewenstein2001}. Traditional optimization falters under such conditions. Climate science has invented or adopted new modes of understanding, such as global model ensembles to explore uncertainty, or hybrid physics-machine learning approaches to tackle more of the system’s complexity \citep{Camps-Valls2025}. Further, proposed approaches embrace the interplay of predictivity and response and tools to support decision strategies that are robust to important classes of deep uncertainty and ignorance. Rather than generating one optimal action recommendation, these solutions explore a set of plausible actions going forward, and iteratively generate new solutions along the way; they can therefore provide acceptable outcomes under a broad range of possible future states of the world \citep{Lempert2006, Shepherd2018, Anderies2024}.

\subsection*{Contextuality$|$Generality: Crossing levels of description}
Complex risk science must continually negotiate between contextuality--the situated, lived, and contingent--and generality--the abstract, systemic, and universal. Too much generality, and we miss the texture of lived reality; too much contextuality, and we lose the ability to act at scale. 

\added{This contextuality$|$generality duality constitutes} our final paradox\deleted{therefore addresses a  central duality of complex risk science: contextuality$|$generality }\added{, embodied by the} risk estimation in the affected communities during the Eaton and Palisades fires\deleted{ embodied this paradox}. The disaster was defined by tensions between generalized wildfire hazard frameworks and locally specific conditions, including neighborhood-specific hazard conditions, infrastructure layouts, and community preparedness practices. Generalizable models support scalability, yet without contextual integration at the scale appropriate to a given event they may systematically misestimate risk in particular locations. This tension illustrates a broader challenge for risk science: how to maintain transferable structure while embedding local knowledge as an active component of model development and interpretation.

In describing the conditions after Hurricane Katrina in New Orleans, author and activist Rebecca Solnit described Federal Emergency Management Agency (FEMA) responses that turned away volunteer help and offers of aid from powerful entities like military ships and Amtrak trains under a lack of knowledge \deleted{about}\added{of} on-the-ground conditions coupled with a presumption that it was not safe to enter New Orleans \citep{Solnit2009}. However, those in the city had a lived knowledge of the conditions, special understanding to perhaps bend or break the rules set in place by more large-scale understanding established in less volatile times. 
“In the hours, days, and weeks after Katrina, those with one set of beliefs were responsible for many deaths; those with another saved many lives. Fear fed by rumors and lies and lurking unexamined beliefs about human nature hit New Orleans like a second hurricane.” In moments of crisis, the inability to cross levels of description, from policy abstractions to lived particulars, becomes catastrophic.

Complexity science helps us understand these failures as problems of transferability of knowledge across location and across scale. All complex systems require ways of translating between contexts, moving more seamlessly from the local to the global, the detailed to the coarse-grained \added{\citep{Simon1991}}. Complexity science is about hierarchical systems, levels of explanation, and coarse-graining. Coarse-graining (a crude look at the whole \citep{Gell-Mann2013}) seeks reduced features, effective variables, or characteristics that can still faithfully describe the system \added{\citep{Rauh2017}}. At the other end of the spectrum, one attempts to put everything of potential consequence in.

The dialectic does not merely move focus to the space between full context and universality, it reveals that understanding the relationship between the scales is what matters. If it is not \textit{either} context \textit{or} generality, then it is a problem of when to have context and when to be general and knowing the opportune times to switch between them.

Resilient systems know how to shift between contextual specificity and generalized response. They neither fixate on local noise nor abstract away the local signal\footnote{\added{Balancing such a tradeoff is well known to the statistics and machine learning communities as the bias–variance tradeoff (see~\citet{Geman1992}). A model with high
variance fits the specific noise of its training data too closely ("fixate on local noise") and fails to generalize; a model with high bias oversimplifies and misses real structure ("abstract away the local signal"). For resilient systems, the optimal mode is not the one that eliminates either error or selects a \deleted{singular}\added{single} mode of operation, but one that manages the tradeoff as conditions change \deleted{(more data, different noise structure)}, which is a formal, quantifiable version of the idea of knowing when and how to move between modes.}}. Their genius lies in knowing when and how to \deleted{move}\added{shift} between modes based on external conditions and internal capacities). \added{This is a flexibility described by \citet{GundersonHolling2002} as `panarchy', which models resilient systems as nested adaptive cycles across scales, where cross-scale movement (not fixation at any one level) is what allows a system to reorganize after disturbance rather than collapse. In individual cognition/human decision-making, too, resilience is not either rule-following (pure generality) or improvising from scratch each time (pure context-fixation), it is a fast alternation between pattern-matching to the general case and testing it against the specific one \citep{Klein1998}. \citet{Cumming2006} bring what we call a system's genius to socio-ecological systems in governance terms: mismatches occur when the scale of an ecological process and the scale of the institution managing it do not align, and resilience specifically requires the capacity to operate, and shift, across scales rather than lock into one.}
\added{This fits the framework of polycentric governance, which argues that complex commons are better served by multiple, overlapping centers of decision-making authority operating at different scales simultaneously, rather than a single institution locked to a single scale~\citep{ostrom_beyond_2010}.}

\added{The ability to switch modes sustains adaptive capacity over time rather than letting it erode \citep{Woods2018}. Networks that sustain adaptability continue to reframe as the sources of variation around them change; by contrast, initially successful adaptations can unwind over time--growing stale, working at cross purposes, failing to keep pace with change--until they suffer sudden performance collapse \citep{Woods2018}. This is precisely the failure a purely general or purely contextual risk science risks: a model tuned to yesterday's conditions that has stopped reframing.} For instance, the body does not adopt one way of fighting off a disease or other external perturbation to the system. Different parts of the body respond in different ways and those ways change across the arc of the disease. The same is true for communities responding to natural hazards--the organization in response must evolve as the disaster plays out. These relationships between complex adaptive systems, resilience, and adaptive capacity open new space for how we must think about and respond to complex risk.

As an example from climate science, growing spatial and temporal connectivity of hydrological cycles drive the acceleration of fast and slow variables leading to destabilisation, tipping points, and new risks related to synchronisation of cycles. Widening interconnections in the water system require a concomitant widening of boundaries typically considered in the study and governance of water. Hydrology reveals how complexity manifests: water cycle extremes are synchronized, tipping points emerge from cross-basin interactions, and governance scales rarely align with ecological ones. As Moore et al. \citep{Moore2024} argue, the question is no longer just fitting to biophysical boundaries but the fitness of governance institutions through flexibility, responsiveness and anticipatory capacity. This fitness is inherently contextual, yet is situated within a global connectivity that demands generalization. This highlights the tension between general system understanding and adaptive institutions.

\section{An Emergent Theory of Change}
\epigraph{There is no end to history, no state of rest for democracy.}{\textit{Danielle Allen} \citep{Allen2023}}
Democracy might be the best system we have come up with for a plural approach to governing our complex societies toward flourishing. Perhaps then it can also guide how we imagine change in (risk) science. In discussions of flourishing, inevitably, the question arises of what to do when different conceptions of a flourishing life come into conflict. What to do when one individual's or group's idea of flourishing or different demands on resources disagree with another's? \added{As introduced above, flourishing is a negotiation between individual meaning and collective well-being; d}\deleted{D}emocracy is the form of governance that we have to support working within this dilemma \citep{Allen2004}. Democracy is not a fixed outcome but an ongoing process of negotiation, repair, and reinvention. Democracy's tenets are dialogue and plurality, both predicated on equality, freedom, and solidarity \added{\citep{Tronto2013}}.

The work before us is to apply these lessons to risk science, and through risk science, to science at large. A plural, dialectical\deleted{, conversational} paradigm is a framework for resilience\deleted{, it is}\added{as well as} \deleted{also} a guide to a theory of change for flourishing. It calls us to reimagine scientific practice, community, and purpose in ways that are open, responsive, and imaginative--capable of sustaining flourishing \added{SETS}\deleted{socio-ecological-technological systems}.

\textit{But how ought scientists and denizens of the world conduct the crisis discipline of risk science?
}

\added{The answer flourishing gives is not to control outcomes, but to give rise to conditions. This is the shift \citet{Nussbaum2011} made in development economics, measuring capability--what a person is able to do and be--in place of achieved outcomes like GDP or attained states. Adaptive capacity performs the same move for systems: it asks how far a system stands from failure, its reserves and its capacity to reconfigure, rather than whether failure has already occurred \citep{Woods2018, BrownWestaway2011}. Democracy performs it again at the level of governance, establishing the conditions--equality, freedom, dialogue--under which differing, even conflicting, conceptions of flourishing can be pursued together, without specifying in advance which conception should win out. Across persons, systems, and polities, flourishing lives in sustained conditions rather than secured outcomes.}

\added{Feedback loops are where this shift from controls to conditions becomes concrete for a system.} The basic operating unit of a system is the feedback loop \citep{Meadows2008}. Feedbacks are the processes by which system change actually occurs\added{, not by dictating an output directly, but by altering the conditions--the loop itself--from which outputs emerge}. Thus, to establish a new paradigm requires identification of the feedbacks that will give rise to change; the dispositions of inquiry and daily activities that will build into systemic transformation. Building on the core ideals of democratic governance we identify \deleted{six}\added{five} seeds (summarized in Fig. \ref{fig:six_principles}), guides to researchers and participants in the planetary community that will create feedbacks leading to the dialectical paradigm:

\subsection*{Interdependence over independence}
Independence is a way of doing science that treats systems and causes and effects as separable, processes as linear, and knowledge as isolated within disciplines. We must acknowledge the interconnectivity between our systems and our sciences and adopt methods that can better represent their connections and interactions. Interdependence is about thinking and acting relationally: designing studies, institutions, and interventions that expect feedback, coevolution, and mutual influence. Interdependence is not a constraint; it’s the condition that makes resilience and flourishing possible.

\subsection*{Responsivity over predictivity}
We must identify methods that permit individuals and communities to create practicable knowledge amongst irreducible extreme uncertainty. The insights from deep uncertainty reframe risk modelling as a practice of navigational orientation rather than deterministic prediction. Instead of asking what will happen, we ask what could plausibly happen, how systems might respond, and what forms of capacity and flexibility we can build. We must develop not merely predictive technologies (which ever more rapidly become inadequate), but ‘response technologies’ (which embody characteristics of resilient systems and actively involve the social element). The models of the future will not simply extend past observations — they will learn, evolve, and participate in shaping the very systems they are meant to describe. The shift is from predictive certainty to adaptive preparedness.

\subsection*{Participation over hegemony}
The prevailing tendency in resilience and risk research has been toward overgeneralization—treating insights born of particular contexts as universal laws. This is not only an epistemic error but a social one, as it privileges detached expertise over lived knowledge. Recognizing risk as volatility and resilience as contextual, participatory ways of knowing acknowledge the irreducible particularity of place, practice, and experience. Context is not a static background but a living tissue of relations continually adapted to changing ecological and social conditions \citep{Scott1998}. That means science should involve, respectfully and with permission, working with communities to understand what solutions to create and how to deploy them. The focus should be on protecting populations and infrastructure rather than prescribing solutions. A participatory science also requires shared languages. The products of risk science–data, models, and visualizations–should act as boundary objects between scientists, decision-makers, and affected communities. Narratives, scenarios, storylines, and imaginaries with quantified uncertainties are an important qualitative improvement over the easily misunderstood probabilistic hazard intensity calculations typical of risk analyses and are more amenable to supporting relationships and meaningful discussions.

\subsection*{Pluralism over integration}
In the study and governance of complex risk, “integration” has often been treated as an ideal–the synthesis of knowledge into a unified framework. Yet for complex systems, integration can be counterproductive. It tends to homogenize perspectives, erasing the very heterogeneity and interaction that give rise to emergence, adaptation, and learning \citep{Olsson2015}. 
\added{This is the seed most directly downstream of dialectic itself: to sustain difference in relation, rather than resolve it into a single frame.} A pluralistic approach, by contrast, values the coexistence of multiple methods, representations, and forms of participation. This is not pluralism as mere inclusion, but as an active process in which diverse ways of knowing encounter one another and generate new understanding. Such pluralism sustains difference long enough for generativity to occur, allowing relationships, meanings, and responses to evolve in context rather than being forced into premature coherence. In this way, pluralism becomes a methodological stance essential to flourishing systems: open, adaptive, and continuously co-creative.

\subsection*{Ongoingness over stationarity}
The great realization of resilience is that it cannot be codified, but must exist and be enacted in infinite variability and context. This emphasis on ‘enaction’ \citep{Varela1991} moves the focus from thing to process, from state to becoming. As much as resilience, and flourishing, are contextual, they are also non-stationary. Thus, they require ongoing engagement and for such unsettledness of concept and solution to be sustainable we require more gradual, agile, and nimble responses. Gradual because there are important transient dynamics over long time scales in addition to critical transitions that need to be responded to \citep{Hastings2018} and agile because shocks are inevitable \citep{Woods2018}. The movement toward ongoingness defines the dialectical paradigm.

\deleted{\textbf{Openness over closedness}
Just as scientific disciplines tend toward hyper-specialization and become siloed, so too do our conversations. We tend toward insularity. Additionally, in the western world our economic incentives promote closing off and making our work proprietary. However, cooperation inherently depends on openness, and collective action on radical openness. Evidence for the value of transparency, sharing, and invitation abounds; from complexity science research where communication between agents at all levels supports better outcomes (see game theory \citep{Gintis2003}) to analyses of the state and progress of science itself \citep{Fortun2017}. Movements like Open Science \citep{Nielsen2020, Gentemann2022}, social justice \citep{Benjamin2022}, and even reform of our media landscape \citep{NewPublic2020} offer guidance for how to move toward more openness and are documenting its manifold impact on these diverse areas of our world. It will be no surprise that they are vital to risk science as well.}

\section{A Civilizational Call}
\epigraph{Individual acts of care or risk-taking can increase solidarity and civic engagement, but they are most likely to do so when embedded in organizations and movements whose governance arrangements facilitate collective actions on behalf of strangers.}{\textit{Margaret Levi} \citep{Levi2023}}

If the dialectical paradigm is the system and the seeds are its \deleted{practices}\added{dispositions}, then their purpose is not only to improve science, but to reorient civilization toward a different relationship with risk--one grounded in mutuality, care, and the possibility of collective flourishing.

The barriers are not only technical but civilizational: they lie in our institutions, our incentives, and our imagination of what science is for. A science for complex risk must remake all three.

The key research challenges ahead are already visible: 
\begin{itemize}
    \item Risk models must embrace non-ergodicity and open-endedness;
    \item Representations must privilege schema and emergent structures over closed forms;
    \item Communicating uncertainty for guiding decisions must turn to storylines, narratives, and scenarios; and 
    \item \added{Theory and} \deleted{M}\added{m}odelling must aim not at finality but at responsiveness and reflexivity.
\end{itemize}

Addressing complex risk requires not only new models but new organizational forms for knowledge production. Traditional disciplinary structures remain essential but are often insufficient for sustained integration across the social, ecological, and technical domains. We suggest that emerging forms of knowledge commons--shared infrastructures for data, models, interpretation, and decision support and governance processes around which communities cohere--can support \textit{research collectives} oriented toward long-term collaboration across institutional and disciplinary boundaries \citep{Hess2011, Bak-Coleman2021}. Such collectives emphasize participation, openness, and shared stewardship of knowledge resources, enabling risk science to better respond to evolving societal needs.

The cascading wildfire example discussed throughout this paper also illustrates the organizational challenge: no single institution owns the coupled knowledge required to characterize infrastructure ignition risk, atmospheric extremes, compound hazards, and community vulnerability simultaneously. Complex risk therefore is not only a modeling problem but a coordination problem, reinforcing the need for knowledge commons–based research structures that support both.

But why can’t science evolve to match this understanding of complex risk science? To translate the dialectic and seeds into a flourishing world, a call must be made to those shaping the structures and incentives of science: \textbf{We must create collectives toward complex risk science, support them with skillfully governed commons, and center care in how we enact our research and evaluation}.

\subsection*{Collectives}
An epistemic posture in most of the seeds above is to value collectivity over individuality. In short, we must create collectives. In science, flourishing means both the well-being of scientists and the production of knowledge that contributes to human and more-than-human well-being. The same is true in society. Flourishing is always mutual: one’s own depends on that of community and environment \citep{deBeauvoir1948}. The interdependence at the heart of complex risk requires that we create more collective structures. 
A collective is not just a group, but a group that reflexively stewards an idea for a time in a way that generates new intelligence, capacity, and meaning. \added{Each element of this definition is worth holding on its own. `Group' names a collection, somehow interacting--but interaction alone is not yet a collective. `Reflexively' names the difference: a collective is in conversation, self-referential, aware of itself as it acts. It `stewards,' actively and on an ongoing basis, rather than simply holding. What it stewards is an `idea,' conceived capaciously--an orientation rather than a possession--which displaces any single person from the center of the collective, without displacing the individual from flourishing itself: flourishing remains, as we have described it, a negotiation of individual meaning and collective well-being, but a collective is organized around what it is oriented toward, not around who leads it. It does this `for a time'--transient, yet ongoing for as long as the orientation holds. And it does so in a way that generates new intelligence, capacity, and meaning--the bridge to what complexity science is teaching us about collective intelligence, and to what may be an emerging sub-discipline in its own right, for how we organize, govern, and relate to one another.}

\added{These elements connect the collective directly to flourishing. Collectives are living systems that generate new capacities and meaning, emerging and reconfiguring with context; they resist the hardening of ideology, remaining attuned to feedback rather than insensitive to changing conditions. Where ideology sacrifices plurality, collapses dialectic, and forecloses conversation, flourishing reveres plurality, works through dialectic, and is itself a conversation. Ideology is fixed; flourishing is ongoing. The collective, in this sense, understands dialectic to be a key dynamic of flourishing--and reveals community itself to be a creative act. The purpose of a collective, in the end, is to expand what everyone in it has access to, and their sense of what they have access to.}

Creating collectives first means understanding how we relate to one another. We must create scientific structures that are welcoming, facilitated, and capable of coexisting with difference. The interactions within them must produce emergent capacity, a potential to be more than the sum of their parts. 
Insights from collective intelligence and the social sciences reveal the conditions that enable such generativity\added{. Collective intelligence is not a constant background hum but an emergent property, arising at particular moments, often catalyzed by an external pressure or a shared goal, rather than existing wherever people are simply co-located \citep{Pentland2015}. Where it does emerge, its mechanism is recognizable across domains: experimentation, communication, selection among candidate solutions, and propagation of what works \citep{Pentland2015}. These domains} \deleted{and} call for new literacies for the complex risk scientist, especially team science, facilitation, and conflict transformation. \added{One among Ostrom's design principles for enduring collective governance is instructive: those affected by a rule must be able to participate in setting and revising it \citep{Ostrom1990}.} These collectives may look very different from today’s research groups, crossing traditional boundaries (institutions, disciplines, sectors of society), tackling wicked problems, organizing and interacting differently, and being oriented toward flourishing.

Indeed, the boundaries of collectives capable of complex risk science for a flourishing world will no longer be held within any single academic institution nor within academia at all. We need institutions and scientific practices that are responsive to each other and to the communities they serve. A global movement for engaged research is growing alongside better understanding of how and when to use participatory approaches as well as how to evaluate their impact on decision-making \citep{Bednarek2022, Wilmer2021, Louder2021, Lemos2018}. Innovative systems of governance that support participatory science, sustained experimentation, and learning may be more capable of managing evolving complex risks \citep{Mach2024}. Study designs can better embrace the challenges and opportunities that humanity brings to real-world experimentation \citep{Wong-Parodi2024}. Boundary organizations, including `informal’ (meaning less traditional, more adaptive, perhaps transient) institutions, can more flexibly and more locally connect to communities and facilitate the creation and transfer of knowledge between science, policymakers, and the public \citep{Collier2009, Goodrich2020}. We need systems and institutions that support and reward this experimentation and interaction, acknowledging the “inefficiency” of doing science within and alongside communities as a vital part of doing it well \citep{Stengers2018}.

\added{Astra Taylor describes civic organizations that come together not through the thin mechanism of a shared petition, but through some cohering energy--a belief, a value, an external driving force--that makes them, at least temporarily, something more than a collection \citep{Taylor2019}. We do not claim these dynamics are unique to risk science. We claim that risk science is a domain in which they cohere and come into sharp focus: the stakes of complex risk make the difference between a collection and a collective consequential in a way few other fields can match.}


\subsection*{Commons}
Collectives are created and sustained through shared commons. Nobel Laureate Elinor Ostrom, in her groundbreaking work on natural resource commons, debunked a decades old myth about the fate of natural resource commons proving that they were not, as believed to be, destined to a tragic abuse and overuse \citep{Ostrom1990}. Late in her career, she turned her attention to a new commons: knowledge. The state of our knowledge commons will determine whether collectives thrive or collapse. Rather than propose ever more sophisticated digital twins or opaque models, we need novel socio-ecological-technological commons that involve not only observational data and uncertainty quantification, but also a participatory ecosystem whereby the whole community maintains, evolves, and governs the shared space. This emphasis reflects growing understanding in the climate services space that more and better data don't necessarily lead to better decisions \citep{Findlater2021}.

\added{A commons, however, does not sustain itself on goodwill alone. Ostrom's design principles hold that a commons must be monitored: once rules are set, a community needs a way of verifying that they are being kept, so that governance rests on accountability rather than solely good faith \citep{Ostrom1990}. For a knowledge commons, monitoring of this kind depends on transparency, sharing, and the usability of what the commons holds--in other words, on open science. We take open science to be transparent and accessible knowledge that is shared and developed through collaborative networks \citep{VICENTESAEZ2018428}, oriented toward rigor, accountability, and reproducibility, and grounded in principles of inclusion, fairness, and equity \citep{Nielsen2020, Gentemann2022}. Open science, in this light, is not one more generic virtue applicable to any field; it is the specific set of practices by which a knowledge commons for complex risk can be built and kept legible to the community that depends on it.}

\added{Most information infrastructures today fall short of this ideal, remaining restrictive either in the tools and know-how required to use them, or directly in cost, as when knowledge and journals sit behind paywalls. The result is a fragmentation of available information, an impoverished lens on the systems under study, and imbalances of knowledge that harden into imbalances of power: asymmetries in access to information reach across a community and determine who can even participate in monitoring the commons in the first place \citep{Fortun2017}. This is a microcosm of the collective action challenge our civilizational call is addressed to: a commons is only as strong as a community's capacity to act together on what it reveals, and open science is the precondition for that capacity, not a separate concern from it.}

\subsection*{Care}
Finally, the scientific community and collectives require an ethic of care \citep{Larrabee1993, Levi2023, Bednar2023}. Perhaps the most challenging component of the risk equation is vulnerability. Part of the challenge is its multifaceted nature: vulnerability is composed of structural, economic, institutional, social, and physical dimensions \citep{Fuchs2012}. The susceptibility of a power grid component is profoundly different from that of a human being, yet our analysis must hold both in view\added{--this is the physical|social dialectic we described earlier, arriving here not as an abstract tension to be studied but as a lived asymmetry to be attended to}. Care offers a way to navigate among these dimensions \added{\citep{PuigdelaBellacasa2017-PUIMOC}}. It includes everything we do to maintain and repair our world--our bodies, our selves, and our environment--so that we can live in it as well as possible \citep{Tronto1990}. Care thus becomes the meeting ground of perspectives on vulnerability and what orients them collectively toward flourishing. \added{Like dialectic and flourishing themselves, care names no terminal state: repair is never finished, and vulnerability never fully resolved, only continually attended to.}

A science of care suggests we need new measures and metrics. Just as GDP fails to capture meaningful lives and futures, so do traditional metrics, including scientific ones, fail to capture the vulnerability and health of communities. Quantitative social science has begun to develop new measures of flourishing \citep{VanderWeele2025b}, but these need to be complemented by insights from political philosophy and social justice \citep{Allen2023, Eliassi-Rad2020}. Reimagining political economy around care \citep{Levi2023, Bednar2023} offers a model that can guide how we assess the flourishing of science and scientists, drawing also on human development frameworks like the capabilities approach \citep{Nussbaum2011}.

\section{Toward a reflexive \added{and connective} scien\deleted{ce}\added{tist}}

Policy-makers and funders must affect the incentives that generate the kind of science that is practiced \citep{Arnott2020}. 
These incentives may involve support for \added{shifting scientific credit and authorship from individual mastery toward collective knowing}\deleted{shifting focus from the individual to the collective}, a process which has always begun with a kind of epistemic humility, a self-emptying that makes space for others' knowing. It is perhaps through a recognition of the importance of self-emptying by which a vision of collective flourishing becomes active and actionable.

Scientists occupy a position of trust in a society where trust is fraying \citep{PewResearchCenter2024}. That trust is a form of social capital, and social capital is for using. In a world with changing extremes, growing vulnerabilities, and undeniable interdependence, how will we put our science (of social, ecological, technological, complex systems) in the world and for what purpose?

Perhaps this is the great lesson from \textit{The Risk Society}: that the literacies and practices of reflexivity are paramount \citep{Beck1992}\added{, practiced individually \citep{Ives2023} as much as collectively, a system of any scale turning its scrutiny on itself}. A civilizational call indeed: to become more individually, collectively, fractally reflexive. Scientists must become more adept at attending to subjective experience \citep{Varela1999}. Subjectivity is not opposed to science but essential to it, especially in complex risk settings. We lack robust methods for apprehending experience, yet this is precisely the need for gaining a better understanding of contextuality and meaningfully including communities in scientific work. Scientists’ reflection on their own work as well as the reflection of ‘lay’ people are both tremendous and mostly untapped potentials. Areas within cognitive science such as micro-phenomenology are making progress to make sense of experience for scientific use and could be translated into SETS research in myriad ways \citep{Petitmengin2001}.

To respond to complex risk is not only to manage hazards, but to mature our ways of knowing and being together. The science that meets this moment will not separate the physical from the social, the deterministic from uncertain, the contextual from the general, but \added{will hold them together, dialectically,} weav\deleted{es}\added{ing} them into a new form of collective intelligence, one capable of holding both the world’s fragility and its generative possibility.



%
\section*{Acknowledgments}
The authors would like to thank the organizers of the 2025 Workshop on Critical Transitions in Socio-Ecological Systems in February 2025 for originally putting the authors in conversation and the ongoing series of `Flourishing Salons' convened by R. McGranaghan in cooperation with the National Academy of Sciences for contributing the bridges between flourishing and (risk) science.


\bibliographystyle{unsrt}
\bibliography{references}

\end{document}